# The adjusted Viterbi training for hidden Markov models

JÜRI LEMBER[1] and ALEXEY KOLOYDENKO[2]

[1]*Tartu University, Liivi 2-507, Tartu 50409, Estonia. E-mail: jyril@ut.ee*
[2]*Division of Statistics, University of Nottingham, Nottingham NG7 2RD, UK.
E-mail: alexey.koloydenko@nottingham.ac.uk*

The *EM procedure* is a principal tool for parameter estimation in the hidden Markov models. However, applications replace EM by *Viterbi extraction*, or *training* (VT). VT is computationally less intensive, more stable and has more of an intuitive appeal, but VT estimation is biased and does not satisfy the following *fixed point property*. Hypothetically, given an infinitely large sample and initialized to the true parameters, VT will generally move away from the initial values. We propose *adjusted Viterbi training* (VA), a new method to restore the fixed point property and thus alleviate the overall imprecision of the VT estimators, while preserving the computational advantages of the baseline VT algorithm. Simulations elsewhere have shown that VA appreciably improves the precision of estimation in both the special case of mixture models and more general HMMs. However, being entirely analytic, the VA correction relies on infinite Viterbi alignments and associated limiting probability distributions. While explicit in the mixture case, the existence of these limiting measures is not obvious for more general HMMs. This paper proves that under certain mild conditions, the required limiting distributions for general HMMs do exist.

## 1. Introduction

Hidden Markov models (HMMs) have been called "one of the most successful statistical modelling ideas that have [emerged] in the last forty years" [8]. Since their classical application to digital communication in 1960s (see further references in [8]), HMMs have had a defining impact on the mainstream technologies of speech recognition [18, 19, 20, 32, 35, 38, 40, 41, 46, 47, 48] and, more recently, bioinformatics [11, 12, 25]. Natural language [21, 36], image [30] and more general spatial [17] models are only a few of the numerous other applications of HMMs.

Applications of HMMs inevitably face the problem of parameter estimation. Let us consider estimation of parameters of a finite-state hidden Markov model (HMM) given observations $x_{1:n} = x_1, \ldots, x_n$ on $X_{1:\infty} = X_1, X_2, \ldots$, the observable process of the HMM, up to time $n$. For any real application, $X_i$ can be assumed to take on values in $\mathcal{X} = \mathbb{R}^D$ for some suitable $D$. Let $Y_{1:\infty} = Y_1, Y_2, \ldots,$ the hidden layer of the HMM, be a (time-homogeneous) Markov chain (MC) with state space $S = \{1, \ldots, K\}$, transition matrix $\mathbb{P} = (p_{ij})$ and initial distribution $\pi = \pi\mathbb{P}$. To every state $l \in S$, there corresponds an emission distribution $P_l(\theta_l)$ with density $f_l$ that is known up to the parametrization $f_l(x; \theta_l)$, $\theta_l \in \Theta_l$, where $\Theta_l$ are rather general domains in $\mathbb{R}^d$. When $Y_k$, $k \geq 1$, is in state $l$, an observation $x_k$ on $X_k$ is emitted according to $P_l(\theta_l)$ and independent of everything else. The $Y_{1:\infty}$ process is also called a *regime* [31]. The *maximum likelihood* (ML) approach has become standard for estimation of $\psi = (\mathbb{P}, \theta)$, the HMM parameters, where $\theta = (\theta_1, \theta_2, \ldots, \theta_K)$. In part, this has been due to the well-known theoretical properties of (local) *consistency* and *asymptotic normality* generally enjoyed by the ML estimators (MLE). Perhaps a more significant reason for the widespread use of the ML approach has been the availability of the EM algorithm with its computationally efficient implementation known as the *Baum–Welch* or simply *Baum*, or *forward–backward algorithm* [1, 2, 8, 14, 20, 39, 40].

Since EM can, in practice, be slow or computationally expensive, it is commonly replaced by *Viterbi extraction*, or *training* (VT), also known as the *Baum–Viterbi* algorithm. VT appears to have been introduced in [19] by F. Jelinek and his colleagues at IBM in the context of speech recognition, in which it has been used extensively ever since [14, 18, 32, 35, 40, 41, 46, 47, 48]. Its computational stability (i.e., rapid exit) and intuitive appeal [14] have also made VT popular in natural language modeling [36], image analysis [30] and bioinformatics [4, 11, 13, 25, 37]. VT is also related to constrained vector quantization [10]. The main idea of the method is to replace the computationally costly expectation (E-step) of the EM algorithm with an appropriate maximization step that generally requires less intensive computer operations (otherwise, the two algorithms scale as $K^2 n$). In speech recognition, essentially the same training procedure was also described by L. Rabiner *et al.* in [22, 41] (see also [39, 40]) as a variation of the *Lloyd algorithm* used in vector quantization. In that context, VT has gained the name *segmental K-means* [14, 22]. The analogy with vector quantization is especially pronounced when the underlying chain is trivialized to i.i.d. variables, thus producing an i.i.d. sample from a mixture distribution. For such mixture models, VT was also described by R. Gray *et al.* in [10], where the training algorithm was considered in the vector quantization context under the name *entropy constrained vector quantization (ECVQ)*. A better-known name for VT in the mixture case is *classification EM (CEM)* [9, 15], stressing that instead of the mixture likelihood, CEM maximizes the *classification likelihood* [4, 9, 15, 33]. VT-CEM was also particularly suitable for the early efforts in image segmentation [44, 45]. Also, for the uniform mixture of Gaussians with a common covariance matrix of the form $\sigma^2 I$ (where $I$ is the $K \times K$ identity matrix) and unknown $\sigma$, VT, or CEM, is equivalent to the *k-means clustering* [9, 10, 15, 43].

### 1.1. VT estimation and relevance of VA to real applications

The VT algorithm for estimation of $\psi$ can be described as follows. Start with some initial values $\psi^{(0)} = (\mathbb{P}^{(0)}, \theta^{(0)})$ and (use the Viterbi algorithm to) find a realization of $Y_{1:n}$ that maximizes the likelihood of the given observations. Any such $n$-tuple of states is called a *Viterbi*, or *forced, alignment*. An alignment partitions the original sample $x_{1:n}$ into subsamples corresponding to distinct states. If regarded as an i.i.d. sample from $P_l(\theta_l)$, the subsample corresponding to state $l$ gives rise to $\hat{\mu}_l^n$, the maximum likelihood estimate (MLE) of $\theta_l$, $l \in S$. At step $m+1$, these estimates replace $\theta^{(m)}$. The transition probabilities are similarly estimated (by MLE) from the current alignment. The updated parameters $\psi^{(m+1)}$ are subsequently used to obtain a new alignment, and so on. It can be shown that, in general, $\psi^{(m)}$ converges (to some $\psi^*(x_{1:n}, \psi^{(0)})$) in finitely many steps $m$ [22]; also, it is usually much faster than the Baum algorithm. Note that when each $f_l$ is modelled as a mixture, which is common in audio and visual processing, VT can be applied at both stages of this model – first, in its general form (i.e., as with $f_l$ general) and then in its CEM form to learn each individual $f_l$. Alternatively, the original HMM can, from the very beginning, be replaced by the equivalent one with hidden states $(l, s(l))$, where $s(l)$ indicates the (sub)component of $f_l$. VT can then also be applied to this new HMM as, for example, has been done in the Philips speech recognition system [35].

Despite its attractiveness, VT can be challenged, as its estimators are generally biased and not consistent. This has been noted, at least in the case of mixtures, since [4], with a specific caveat issued in [49]. Simulations in [27] and [24] illustrate appreciable biases in VT estimation in the i.i.d. and more general HMM settings, respectively. At the same time, these facts are not surprising. Indeed, unlike EM, which increases the likelihood of $\psi$ given $x_{1:n}$, VT increases the joint likelihood of the (hidden) state sequence $y_{1:n}$ and the parameters $\psi$, given $x_{1:n}$. According to [34], under certain conditions, the difference between the two objective functions vanishes as $D$, the dimension of the emission $X_i$, grows sufficiently large relative to $\log(K)$, which can be realistic in *isolated word recognition* [34]. However, as later clarified in [14], this does not imply closeness of the parameter estimates obtained by EM and VT (unless the algorithms are initialized identically) since both perform a local, rather than global, optimization.

Certainly, unbiasedness and consistency are neither necessary nor sufficient for a procedure to perform well in applications [45]. However, there are indications that some applications, such as *segment-based speech recognition* [46], do prefer the standard, that is, EM-type, likelihood maximization. Also, [46] notes that *conventional speech recognizers* would prefer the 'smoother convergence' of $\psi^{(m)}$ under EM, presumably over the more abrupt, greedy convergence of $\psi^{(m)}$ under VT. At the same time, it appears that in complex environments, VT can be appreciably simpler to implement than EM [46]. Hence, it appears to be sensible to combine the simplicity of VT's implementation with the desirable properties of EM.

Indeed, there are variations of VT that use more than one best alignment or several perturbations of the best alignment [36]. VA, our type of adjusted VT, is of a different nature as it improves the estimation precision by means of analytic calculations and does not compute more than one optimal alignment per iteration. Moreover, we suggest that investigating such alternatives to VT and EM for real applications is nowadays much more appealing than ever before, thanks to the abundance of virtually infinite and freely available streams of audio and video (e.g., real-time broadcasting) as well as biological data. Actually, practitioners have already realized this by shifting from entirely supervised to semi- and unsupervised modes of training [50]. One naïve realization of these ideas is to simply use the estimates obtained from a labeled sample (i.e., with $y_{1:n}$ known) as the initial guess $\psi^{(0)}$ for a further unsupervised retraining. A more dedicated application would be *model adaptation*, wherein the model $\psi^{(0)}$ (initially trained in any mode) may need to be adapted to a new environment (e.g., speaker) differing from the original one mostly, or only, by the emission parameters. Applicability of our adjusted VT for mixture models and situations when the transition probabilities are either known or nuisance is further discussed in Section 2.3. Finally, simulations in [27] and [24] clearly show that VA, our method of adjusting VT, does significantly improve the precision of VT estimation. In those experiments, the VA estimates are always comparable to the EM estimates, while the VA algorithm is only marginally more intensive than the baseline VT algorithm.

## 1.2. The adjusted Viterbi training and contribution of this work

Is it possible to adjust VT in an analytic way in order to enjoy both the desirable properties of VT (fast convergence of $\psi^{(m)}$, overall computational feasibility, simplicity of implementation and an overall intuitive appeal) and more consistent estimation? Ensuring that an algorithm *has the true parameters as its asymptotically fixed point* turns out to be pivotal in constructing such adjusted estimators. Evidently, this fixed point property holds for EM, but not for VT. Namely, for a sufficiently large sample, the EM algorithm 'recognizes' and 'confirms' the true parameters. In contrast to this, an iteration of VT generally disturbs the correct values noticeably. In [27], we have proposed to modify VT in order to make the true parameters an asymptotically fixed point of VA, the resulting algorithm.

In order to understand VA, it is crucial to understand the asymptotic behaviors of $\hat{\mu}_l^n$ and $\hat{p}_{ij}^n$, the maximum likelihood estimators based on the Viterbi alignment. Since the alignment depends on $\psi^{(0)}$, the initial values of the parameters (and on the tie-breaking rule, which is ignored for the time being), so do $\hat{\mu}_l^n(\psi^{(0)}, X_{1:n})$ and $\hat{p}_{ij}^n(\psi^{(0)}, X_{1:n})$. Note that, for $\psi$ to be asymptotically fixed by an estimation algorithm, it means that if $\psi = (\mathbb{P}, \theta)$ are the true parameters and are used to compute the alignment, then

$$\hat{\mu}_l^n(\psi, X_{1:n}) \underset{n\to\infty}{\longrightarrow} \theta_l \quad \text{a.s. } \forall l \in S; \qquad \hat{p}_{ij}^n(\psi, X_{1:n}) \underset{n\to\infty}{\longrightarrow} p_{ij} \quad \text{a.s. } \forall (i,j) \in S^2. \quad (1.1)$$

The reason why VT does not enjoy the desired fixed point property is that (1.1) need not hold in general [4, 49]. Hence, in order to restore the above fixed point property in VT, we need to verify that the sequences in (1.1) converge almost surely and, provided they do, exhibit their limits. This paper essentially accomplishes these tasks. Namely, we show that (under certain mild conditions) the empirical measures $\hat{P}_l^n(\psi, X_{1:n})$ obtained via the Viterbi alignment do converge weakly to a certain limiting probability measure $Q_l(\psi)$ (2.5) and that, in general, $Q_l(\psi) \neq P_l(\theta_l)$. In [24], we have shown that under general conditions on the densities $f_l(x; \theta_l)$ (and, for $\Theta_l$, closed subsets of $\mathbb{R}^d$), the above convergence $\hat{P}_l^n(\psi, X_{1:n}) \Rightarrow_{n\to\infty} Q_l(\psi)$ a.s. (properly introduced in (2.5)) implies convergence of $\hat{\mu}_l^n$, that is,

$$\hat{\mu}_l^n(\psi, X_{1:n}) \underset{n\to\infty}{\longrightarrow} \mu_l(\psi) \qquad \text{a.s., where } \mu_l(\psi) \stackrel{\text{def}}{=} \arg\max_{\theta_l' \in \Theta_l} \int \ln f_l(x; \theta_l') Q_l(\mathrm{d}x; \psi). \quad (1.2)$$

Since, in general, $Q_l(\psi) \neq P_l(\theta_l)$, clearly $\mu_l(\psi)$ need not equal $\arg\max_{\theta_l'} \int \ln f_l(x; \theta_l') \times P_l(\mathrm{d}x; \theta_l)$.

In order to obtain the above results, in Section 4, we extend Viterbi alignments, or paths, ad infinitum. Namely, considering (finite) Viterbi alignments with tie-breaking rules of a special kind, we prove the existence of a decoding $v : \mathcal{X}^\infty \to S^\infty$ such that, for almost every realization $x_{1:\infty}$, the following property holds: for every $m \in \mathbb{N}$, there exists an $n = n(x_{1:\infty}, m) \in \mathbb{N}$, $n > m$, such that the codeword $v(x_{1:\infty})$ and the Viterbi alignment based on $x_{1:n}$ agree up to time $m$. To emphasize the dependence of $v$ on $\psi$, we will write $v(x_{1:\infty}; \psi)$. It can then also be shown that when $\psi$ are the true parameters, the process $V \stackrel{\text{def}}{=} v(X_{1:n}; \psi)$ is regenerative. In particular, for any $i, j \in S$, there exists $q_{ij}(\psi) \geq 0$ such that $\sum_j q_{ij}(\psi) = 1$ for every $i \in S$ and

$$\hat{p}_{ij}^n(\psi; X_{1:n}) \xrightarrow[n\to\infty]{\text{a.s.}} q_{ij}(\psi). \quad (1.3)$$

Again, in general, $p_{ij} \neq q_{ij}(\psi)$. Reduction of the biases $\mu_l(\psi) - \theta_l$ and $q_{ij}(\psi) - p_{ij}$ is the main feature of the adjusted Viterbi training.

### 1.3. Previous related work

We are not aware of any systematic treatment of asymptotic reduction of the bias in VT estimation (without compromising the advantages of the VT algorithm over Baum–Welch) preceding [27]. In [23], however, a sequential version of VT ('the segmental K-means algorithm') is suggested, which can allegedly reduce the estimation bias asymptotically. The suggested modification appears substantially different from our adjustment, although we have been unable to evaluate the algorithm of [23] thoroughly due to the lack of detail in its description in [23] or anywhere else to date.

Moreover, to the best of our knowledge, there has been no systematic study of asymptotic properties of the Viterbi alignments to date besides certain attempts made by Kogan in [23] in the context of the sequential version of VT (see above) and, more recently, by Caliebe and Rösler in [7] and Caliebe in [5]. Both groups have given thorough treatments of certain special cases, mostly $K = 2$, but this, as we explain below, is *too* special.

Importantly, it was recognized in [23] that under certain conditions, longer Viterbi alignments can be obtained piecewise. Roughly, the end-points of the pieces and the (random) times of their occurrence were termed 'special columns' and 'most informative stopping times', respectively. In [5, 7], related notions of 'meeting states' and 'meeting times' are used. Independently of [5, 7, 23], we have built our theory on the notion of nodes (roughly, observations emitted from the 'special columns'; see Section 3.1) and the stopping times of their occurrence. If defined to be independent of a particular global tie-breaking rule, the meeting times of [5] would correspond to 'strong nodes' of order 0, a particular type of nodes. More importantly, even our (general) nodes, which are essentially equivalent to the special columns of [23] and 'path crossings' of [5, 7], are *not sufficiently general* in the sense that HMMs with aperiodic and irreducible Markov chains need not necessarily have special columns, or nodes, infinitely often almost surely, despite the claim to the contrary made in Theorem 2 of [23] (stated without proof and implicitly cited in [14]). For a counterexample, we refer to Example 3.11 in [26], a downloadable preprint of this paper. Appropriate sufficient conditions to guarantee the desired property have also been given in [26] for the first time. Implicitly, the alignment process in [23] was recognized as regenerative with respect to the 'most informative stopping times'. The limiting alignment process of [5] is already explicitly shown to be regenerative. Regenerativity with respect to (the times of) nodes is also essential for our purpose of exhibiting the limiting measures $Q_l(\psi)$ (2.5) and $q_{ij}(\psi)$ (1.3).

Convergence of the Viterbi paths was, to the best of our knowledge, first seriously considered in [5, 7], where the existence of infinite alignments for certain special cases, such as $K = 2$ and some HMMs with additive white Gaussian noise, was also proven. While innovative, the main result of [7] (Theorem 2) makes several restrictive assumptions preventing its extension beyond the $K = 2$ case. As its by-product, this work extends some, and corrects other, results of [5, 7]. This is explained in detail in the appropriate paragraphs of Sections 3.1– 3.3 and Section 4. Also, note that our goal of exhibiting $Q_l(\psi)$ and $q_{ij}(\psi)$ extends beyond solely defining infinite Viterbi alignments (the main goal of [7]).

## 1.4. Organization of the rest of the paper

First, in Section 2, we properly introduce the baseline and adjusted Viterbi training procedures (Section 2.2) for HMMs. In Section 2.3, the adjusted Viterbi training is discussed in the context of the following two important variations on the main situation: the regime parameters are known or nuisance. More general issues of implementation are discussed in Section 2.4. Sections 2.3 and 2.4 can be skipped without interruption of the main presentation.

Recall that our ultimate goal has been asymptotic reduction of the bias in VT estimation for as general a class of HMMs as possible. The main goal of this paper, however, is to prove the existence of the limiting measures $Q_l(\psi)$ (2.5) and $q_{ij}(\psi)$ (1.3) that underpin our approach to achieving the ultimate goal. A significant effort has been made to achieve this accurately and under as non-restrictive conditions as possible. This is the main reason why we cannot directly reuse the tools used by others ([5, 7, 23]). As we reiterate further in Section 3, the asymptotic behavior of the Viterbi alignment is not trivial and does require special tools. Thus, *nodes* and *barriers*, our main tools, are presented in Sections 3.1 and 3.3, respectively. In Section 3.2, we explain our piecewise construction of the *proper* Viterbi alignments. This is still at the level of individual realizations of the HMM process. In Section 3.3, barriers, on the other hand, extend our construction for almost every realization of the HMM process. This is the essence of Lemmas 3.1 and 3.2, the first of the two main results of this paper. In Section 4, we define $V_{1:\infty}$, the *proper infinite alignment process*. Finally, in the same section we prove the existence of the measures $Q_l(\psi)$ and $q_{ij}(\psi)$, our second main result, using regenerativity of the augmented process $(V_{1:\infty}, X_{1:\infty})$ (Theorem 4.1 and Corollary 4.1).

Exhibiting the measures $Q_l(\psi)$ under very general conditions has necessitated several rather technical constructions, mainly used to prove Lemmas 3.1 and 3.2. Due to spatial limitations, they are not given here, but rather appear in [26].

## 2. The adjusted Viterbi training

### 2.1. The model

Recall that $Y_{1:\infty}$ takes values in $S = \{1, \ldots, K\}$ and has transition matrix $\mathbb{P}$. Let $Y_{1:\infty}$ be irreducible and aperiodic, hence a unique $\pi = \pi \mathbb{P}$ exists. Let the *emission distributions* $P_l(\theta_l)$, $l \in S$, be defined on $(\mathcal{X}, \mathcal{B})$, where $\mathcal{X}$ and $\mathcal{B}$ are a separable metric space and the corresponding Borel $\sigma$-algebra, respectively. Let $f_l$ be the density of $P_l(\theta_l)$ with respect to a suitable reference measure $\lambda$ on $(\mathcal{X}, \mathcal{B})$.

**Definition 2.1.** *The stochastic process $X$ is a hidden Markov model if there is a (measurable) function $h$ such that, for each $n$,*

$$X_n = h(Y_n, e_n), \qquad \text{where } e_1, e_2, \ldots \text{ are i.i.d. and independent of } Y. \qquad (2.1)$$

Hence, the emission distribution $P_l(\theta_l)$ is the distribution of $h(l, e_n)$. The distribution of $X$ is completely determined by the regime parameters $\mathbb{P}$ and the emission distributions $P_l(\theta_l)$, $l \in S$. The process $X$ is also $\alpha$-mixing and, therefore, ergodic [14, 16, 29].

### 2.2. Viterbi alignment and training

Let

$$\Lambda(y_{1:n}; x_{1:n}, \psi) = \mathbf{P}(Y_{1:n} = y_{1:n}) \prod_{i=1}^{n} f_{y_i}(x_i; \theta_{y_i}), \qquad \text{where } \mathbf{P}(Y_{1:n} = y_{1:n}) = \pi_{y_1} \prod_{i=2}^{n} p_{y_{i-1} y_i},$$

be the likelihood functions of the $y_{1:n}$, treated as parameters. Given $x_{1:n}$, let $\mathcal{V}(x_{1:n}; \psi)$ be the set of all maximum-likelihood estimates of $y_{1:n}$. These estimates, or paths, are efficiently obtained by the Viterbi algorithm and are called the *Viterbi alignments*.

The non-uniqueness of the alignments causes substantial technical inconveniences. In Section 3.2, we specify unique $v(x_{1:n}; \psi) \in \mathcal{V}(x_{1:n}; \psi)$ for every $n \in \mathbb{N}$ and $x_{1:n} \in \mathcal{X}^n$ (and every $\psi$) in a consistent manner that is suitable to prove the existence of $Q_l(\psi)$. Meanwhile, the uniqueness of $v(x_{1:n}; \psi)$ is an assumption. VT estimation of $\psi$ is defined formally as follows (where $I_A$ is the indicator function of set $A$):

(1) choose initial values for the parameters $\psi^{(k)} = (\mathbb{P}^{(k)}, \theta^{(k)})$, $k = 0$;
(2) given $\psi^{(k)}$, current parameters, obtain the alignment $v^{(k)} = v(x_{1:n}; \psi^{(k)})$;
(3) update the regime parameters $\mathbb{P}^{(k+1)} \stackrel{\text{def}}{=} (\hat{p}_{ij}^n)$ as given by

$$\hat{p}_{ij}^n \stackrel{\text{def}}{=} \begin{cases} \dfrac{\sum_{m=1}^{n-1} I_{\{i\}}(v_m^{(k)}) I_{\{j\}}(v_{m+1}^{(k)})}{\sum_{m=1}^{n-1} I_{\{i\}}(v_m^{(k)})}, & \text{if } \sum_{m=1}^{n-1} I_{\{i\}}(v_m^{(k)}) > 0, \\ \mathbb{P}_{ij}^{(k)}, & \text{otherwise}, \end{cases} \quad i,j \in S; \quad (2.2)$$

(4) assign $x_m$, $m = 1, 2, \ldots, n$, to class $v_m^{(k)}$ and, equivalently, define empirical measures

$$\hat{P}_l^n(A; \psi^{(k)}, x_{1:n}) \stackrel{\text{def}}{=} \frac{\sum_{m=1}^n I_{A \times \{l\}}(x_m, v_m^{(k)})}{\sum_{m=1}^n I_{\{l\}}(v_m^{(k)})}, \quad A \in \mathcal{B}, l \in S; \quad (2.3)$$

(5) for each class $l \in S$, obtain $\hat{\mu}_l^n(\psi^{(k)}, x_{1:n})$, MLE of $\theta_l$, given by

$$\hat{\mu}_l^n(\psi, x_{1:n}) \stackrel{\text{def}}{=} \arg\max_{\theta_l' \in \Theta_l} \int \ln f_l(x; \theta_l') \hat{P}_l^n(\mathrm{d}x; \psi, x_{1:n}) \quad (2.4)$$

and for all $l \in S$, let

$$\theta_l^{(k+1)} \stackrel{\text{def}}{=} \begin{cases} \hat{\mu}_l^n(\psi^{(k)}, x_{1:n}), & \text{if } \sum_{m=1}^K I_{\{l\}}(v(x_{1:n}; \psi^{(k)})_m) > 0, \\ \theta_l^{(k)}, & \text{otherwise}. \end{cases}$$

To better interpret VT, suppose that, at some step $k$, $\psi^{(k)} = \psi$, thus $v^{(k)}$ is obtained using the true parameters. Let $y_{1:n}$ be the actual hidden realization of $Y_{1:n}$. The training 'pretends' that the alignment $v^{(k)}$ is perfect, that is, that $v^{(k)} = y_{1:n}$. If the alignment were indeed perfect, then the empirical measures $\hat{P}_l^n$, $l \in S$, would be obtained from the i.i.d. samples generated from $P_l(\theta_l)$ and the MLE $\hat{\mu}_l^n(\psi, X_{1:n})$ would be natural estimators to use. Under these assumptions, $\hat{P}_l^n(\psi, X_{1:n}) \Rightarrow P_l(\theta_l)$ as $n \to \infty$ a.s. and, provided that $\{f_l(\cdot; \theta_l): \theta_l \in \Theta_l\}$ is a $P_l$-Glivenko–Cantelli class and $\Theta_l$ is equipped with a suitable metric, we would have $\lim_{n\to\infty} \hat{\mu}_l^n(\psi, X_{1:n}) = \theta_l$ a.s. Hence, if $n$ is sufficiently large, then $\hat{P}_l^n(\psi, X_{1:n}) \approx P_l(\theta_l)$ and $\theta_l^{(k+1)} = \hat{\mu}_l^n(\psi, x_{1:n}) \approx \theta_l = \theta_l^{(k)}$ for every $l \in S$. Similarly, if the alignment is perfect, then $\lim_{n\to\infty} \hat{p}_{ij}^n(\psi, X_{1:n}) = \mathbf{P}(Y_2 = j | Y_1 = i) = p_{ij}$, a.s. Thus, for the perfect alignment, $\psi^{(k+1)} = (\mathbb{P}^{(k+1)}, \theta^{(k+1)}) \approx (\mathbb{P}^{(k)}, \theta^{(k)}) = \psi^{(k)} = \psi$, that is, $\psi$ would be (approximately) a fixed point of the training algorithm. Certainly, the alignment, in general, is not perfect, even when it is computed with the true parameters. In particular, the empirical measures $\hat{P}_l^n(\psi, X_{1:n})$ can be rather far from those based on i.i.d. samples from $P_l(\theta_l)$. Hence, we have no reason to expect that $\lim_{n\to\infty} \hat{\mu}_l^n(\psi, X_{1:n}) = \theta_l$ a.s. and $\lim_{n\to\infty} \hat{p}_{ij}^n(\psi, X_{1:n}) = p_{ij}$ a.s. Moreover, we do not even know whether the sequences of empirical measures $\hat{P}_l^n(\psi, X_{1:n})$, or MLE estimators $\hat{\mu}_l^n(\psi, X_{1:n})$ and $\hat{p}_{ij}^n(\psi, X_{1:n})$, converge almost surely at all.

As stated in Theorem 4.1, under certain mild conditions, there exist probability measures $Q_l(\psi)$, $l \in S$, such that

$$\hat{P}_l^n(\psi, X_{1:n}) \underset{n \to \infty}{\Longrightarrow} Q_l(\psi) \qquad \text{a.s.} \tag{2.5}$$

From the proof of Theorem 4.1, it also follows (Corollary 4.1) that for every $i \in S$, there exist probabilities $q_{i1}, \ldots, q_{iK}$ such that (1.3) holds. In general, $\mu_l(\psi) \neq \theta_l$ and $q_{ij}(\psi) \neq p_{ij}$. In order to reduce the biases $\theta_l - \mu_l(\psi)$ and $p_{ij} - q_{ij}(\psi)$, we have proposed the *adjusted Viterbi training*. Namely, suppose that (1.2) and (1.3) hold and consider the mappings

$$\psi \mapsto \mu_l(\psi), \qquad \psi \mapsto q_{ij}(\psi), \qquad l, i, j = 1, \ldots, K. \tag{2.6}$$

The functions in (2.6) do not depend on $x_{1:n}$, hence the following corrections are well defined:

$$\Delta_l(\psi) \stackrel{\text{def}}{=} \theta_l - \mu_l(\psi), \qquad R_{ij}(\psi) \stackrel{\text{def}}{=} p_{ij} - q_{ij}(\psi), \qquad l, i, j = 1, \ldots, K. \tag{2.7}$$

Based on (2.7), the *adjusted Viterbi training* replaces VT steps (3) and (5) as given below:

(3) for every $i, j \in S$, update the matrix $\mathbb{P}^{(k+1)} \stackrel{\text{def}}{=} (p_{ij}^{(k+1)})$ according to

$$p_{ij}^{(k+1)} \stackrel{\text{def}}{=} \hat{p}_{ij}^n + R_{ij}(\psi^{(k)}); \tag{2.8}$$

(5) for all $l \in S$, let

$$\theta_l^{(k+1)} \stackrel{\text{def}}{=} \Delta_l(\psi^{(k)}) + \begin{cases} \hat{\mu}_l^n(\psi^{(k)}, x_{1:n}), & \text{if } \sum_{m=1}^{K} I_{\{l\}}(v_m) > 0, \\ \theta_l^{(k)}, & \text{otherwise.} \end{cases}$$

Provided $n$ is sufficiently large, VA, as desired, has the true parameters $\psi$ as its (approximately) fixed point. Indeed, suppose that $\psi^{(k)} = \psi$. From (1.2), $\hat{\mu}_l^n(\psi^{(k)}, x_{1:n}) = \hat{\mu}_l^n(\psi, x_{1:n}) \approx \mu_l(\psi) = \mu_l(\psi^{(k)})$ for all $l \in S$. Similarly, from (1.3), $\hat{p}_{ij}^n(\psi^{(k)}, x_{1:n}) = \hat{p}_{ij}^n(\psi, x_{1:n}) \approx q_{ij}(\psi) = q_{ij}(\psi^{(k)})$ for all $i, j \in S$. Thus,

$$\theta_l^{(k+1)} = \hat{\mu}_l^n(\psi, x_{1:n}) + \Delta_l(\psi) \approx \mu_l(\psi) + \Delta_l(\psi) = \theta_l = \theta^{(k)}, \qquad l \in S, \tag{2.9}$$

$$p_{ij}^{(k+1)} = \hat{p}_{ij}^n(\psi, x_{1:n}) + R_{ij}(\psi) \approx q_{ij}(\psi) + R_{ij}(\psi) = p_{ij} = p_{ij}^{(k)}, \qquad i, j \in S. \tag{2.10}$$

Hence, $\psi^{(k+1)} = (\mathbb{P}^{(k+1)}, \theta^{(k+1)}) \approx (\mathbb{P}^{(k)}, \theta^{(k)}) = \psi^{(k)}$.

**Example 1 (Mixtures).** Let $X_1, X_2, \ldots$ be i.i.d. and follow a mixture distribution with density $\sum_{l=1}^{K} \pi_l f_l(x; \theta_l)$ and (positive) mixing weights $\pi_l$. Such a sequence is an HMM with transition probabilities $p_{ij} = \pi_j$ for all $i, j \in S$. In this special case, the alignment and the measures $Q_l$ are easy to exhibit. Indeed, for any set of parameters $\psi = (\pi, \theta)$, the alignment $v(x_{1:n}; \psi)$ can be obtained via a generalized *Voronoi partition* $\mathcal{S}(\psi) = \{S_1(\psi), \ldots, S_K(\psi)\}$, where

$$S_1(\psi) = \{x \in \mathcal{X} : \pi_1 f_1(x; \theta_1) \geq \pi_j f_j(x; \theta_j), \forall j \in S\}, \tag{2.11}$$

$$S_l(\psi) = \{x \in \mathcal{X} : \pi_l f_l(x; \theta_l) \geq \pi_j f_j(x; \theta_j), \forall j \in S\} \setminus \bigcup_{k=1}^{l-1} S_k(\psi), \quad l = 2, \ldots, K. \tag{2.12}$$

Now, the alignment can be defined pointwise as follows: $v(x_{1:n}; \psi) = (v(x_1; \psi), \ldots, v(x_n; \psi))$, where $v(x; \psi) = \sum_{k=1}^{K} k I_{S_k(\psi)}(x)$, which returns $l$ if and only if $x \in S_l(\psi)$. The convergence (2.5) now follows immediately from the strong law of large numbers. Indeed, if $\psi$ are the true parameters and if the alignment is obtained based on $\psi$, then the SLLN immediately gives $\hat{P}_l^n(\psi) \Rightarrow Q_l(\psi)$ almost surely, with densities $q_l(x; \psi)$ of $Q_l(\psi) \propto f(x; \psi) I_{S_l(\psi)} = (\sum_{k=1}^{K} \pi_k f_k(x; \theta_k)) I_{S_l(\psi)}$, $l = 1, 2, \ldots, K$. Hence, the limit of the class-conditional MLE $\hat{\mu}_l^n$ is given by

$$\mu_l(\psi) = \arg\max_{\theta_l' \in \Theta_l} \int_{S_l(\psi)} \ln f_l(x; \theta_l') \left(\sum_{k=1}^{K} \pi_k f_k(x; \theta_k)\right) \lambda(\mathrm{d}x), \tag{2.13}$$

which, depending on the model, can differ from $\theta_l$ significantly ([24, 27]). Also, (1.3) follows easily in this case (see [27] for further details). Namely, note that

$$\hat{\pi}_l^n(\psi, X_{1:n}) \xrightarrow[n \to \infty]{\text{a.s.}} q_l(\psi) = \sum_{k=1}^{K} \pi_k \int_{S_l(\psi)} f_k(x; \theta_k) \lambda(\mathrm{d}x). \tag{2.14}$$

Thus, in the special case of mixtures, the adjustments $\Delta_l$ and $R_l$ are relatively easy to obtain and the adjusted Viterbi training is easy to implement. The simulations in [27] have largely supported the theory, demonstrating both the computational advantage of VA over EM and the increased precision of VA relative to VT.

## 2.3. VA for 'independent training'

Some applications, such as large vocabulary speech recognition systems [35], fix the regime parameters exogenously. With the appropriate simplifications, the baseline and adjusted Viterbi training procedures, as well as the EM algorithm, immediately apply in such situations. In fact, in [24, 27], VA is discussed primarily in this simplified context. It can then be argued that, when the regime parameters are known, VA is unnecessary as *MLI*, the maximum likelihood estimation under the independence assumption (which can also be called *independent training*), applies. Let us discuss this issue in more detail. According to [31], MLI estimates the emission parameters (and possibly $\pi$ when $\mathbb{P}$ is unknown and not of interest) of general (ergodic) HMMs pretending that $Y_1, Y_2, \ldots$, are independent, that is, the entire HMM follows a mixture model. This is appealing since the marginal distribution of the emissions of any HMM (with a stationary regime) is indeed the mixture with density $\sum_k \pi_k f_k(\cdot; \theta_k)$. Thus, MLI is an instance of the *maximum pseudo-likelihood* (MPL) based on the above mixture approximation. The MLI–MPL estimators for the emission parameters are (locally) consistent [31, 42] and can also be delivered by EM (for mixtures). Similarly to the general case, when computational resources do matter, VT (for mixtures) can also be used instead of EM in this case. As in the general case, Baum–Welch and VT scale identically, but their common computational complexity is now $Kn$, as opposed to $K^2 n$. The comparative computational performances of Baum–Welch and VT for mixtures and in the general case are also similar (the Baum algorithm involves more intensive operations). At the same time, as Example 1 in Section 2.2 above shows, the VT estimators are still not consistent and, in particular, the correction $\Delta_l = \theta_l - \mu_l(\psi)$, with $\mu_l(\psi)$ as in (2.13), can be significant.

Let us make another point. Let $\theta$ be fixed and let $\Delta_l$ and $\Delta_l^*$ be the corrections obtained with and without the independence assumption ($p_{ij} = \pi_j$, $i, j \in S$), respectively. The following intuitive fact has been shown in [24] by simulation: $\Delta_l^* \leq \Delta_l$ and the difference $\Delta_l - \Delta_l^*$ widens as the dependence in $\mathbb{P}$ becomes stronger. This suggests that there is more to gain by adjusting VT for mixtures toward MPL-MLI than by adjusting VT for the actual HMM toward the true MLE. Thus, if one is interested in a computationally efficient approximation to (the Baum implementation of) MPL–MLI, the adjusted Viterbi training for mixtures is a sensible alternative to the baseline Viterbi training for mixtures. Also, note that VA for mixture models was studied in [27], where, in addition to the theoretical demonstration of the VT bias, it was also shown by simulations that this bias could be *significantly* reduced by VA. Importantly, in the mixture case, the VA corrections are often given explicitly, which simplifies the implementation of the algorithm.

The independent training approach is also a natural choice when the underlying regime is a general ergodic process (not necessarily Markov) with an (unknown) stationary distribution $\pi$. Even when not of direct interest, $\pi$ can and needs to be estimated. Again, if computational efficiency is an issue, VA for mixtures with unknown weights is an alternative to the Baum algorithm (for mixtures with unknown weights). Note that in this case, the corrections $R_l = \pi_l - q_l(\psi)$, with $q_l(\psi)$ as in (2.14), should be used in addition to the $\Delta_l$ corrections. Simulations in [27] showed a clear advantage of using both adjustments $R_l$ and $\Delta_l$ for mixture models with unknown $\pi$. In particular, VA was, as usual, both superior to VT and only slightly inferior to EM, in precision. Remarkably, taking few steps to stabilize, VA also outperformed VT in total runtime.

## 2.4. Implementation

To implement VA in practice, explicit expressions for $Q_l(\psi)$ (or $\mu_l(\psi)$) and $q_{ij}(\psi)$ are desirable. In general, however, these functions can be very difficult to compute with high precision. At the same time, as was just pointed out in Section 2.3 above, the corrections $\Delta_l$ and $R_l$ are easy to obtain for a broad class of mixture models including the most commonly used mixtures of Gaussians with equal and known covariances. Other details of VA implementation have been addressed in [27] and [24] for mixture and more general models, respectively. For one example, [24] discusses the *stochastically adjusted Viterbi training*, an efficient implementation of VA for general HMMs when the corrections cannot be obtained analytically. Although simulations do require extra computations, the overall complexity of the stochastically adjusted VT can still be considerably lower than that of Baum–Welch. Certainly, this requires further investigation. Other practical issues are also a subject of continuing investigation.

## 3. Infinite Viterbi alignment

The idea of the adjusted Viterbi training is based on, firstly, the observation that the maximum likelihood path (the Viterbi alignment) differs substantially from the underlying Markov chain and, secondly, that these differences need to be accounted for in order for the overall HMM-based inference to be accurate. Our adjusted Viterbi training need not be the only method to correct the training process for these differences. However, any such method must inevitably appreciate the asymptotic properties of both the Viterbi alignment and the subsamples of the emissions as classified by the alignment. After all, it is these features that determine the properties of the VT estimators in general and the asymptotic bias of VT in particular.

Even disregarding the non-uniqueness of the Viterbi alignment $v(x_{1:n})$ (dependence on $\psi$ is temporarily suppressed), the asymptotic behavior of $v(X_{1:n})$ is not trivial since the $(n+1)$th observation can in principle change the entire alignment based on $x_{1:n}$. Namely, let $v(x_{1:n})$ and $v(x_{1:n+1})$ be the alignments based on $x_{1:n}$ and $x_{1:n+1}$, respectively. It might happen with positive probability that $v(x_{1:n})_i \neq v(x_{1:n+1})_i$ for every $i = 1, \ldots, n$. At the same time, the fact that the alignment changes infinitely often makes it difficult to define a meaningful infinite alignment process. For most HMMs, however, there is a positive probability of observing $x_{1:n}$ such that, regardless of the value of the $(n+1)$th observation (provided $n$ is sufficiently large), the alignments $v(x_{1:n})$ and $v(x_{1:n+1})$ agree for a sufficiently long time $u \leq n$. Consequently, regardless of what happens in the future, the first $u$ elements of the alignment remain constant. Provided that there is an increasing unbounded sequence $u_i$ ($u < u_1 < u_2 < \cdots$) such that the alignment up to $u_i$ remains constant, infinite alignments can then be defined. The observation that for most commonly used HMMs, a typical realization $x_{1:\infty}$ has infinitely many $u_i$ is the basis of our further analysis.

Consider the following simple model that guarantees almost every $x_{1:\infty}$ to have infinitely many $u_i$'s and provides an insight into a significantly more general scenario. Let state $1 \in S$ and event $A \in \mathcal{B}$ be such that $P_1(A) > 0$, while $P_l(A) = 0$ for $l = 2, \ldots, K$. Thus, any observation $x_u \in A$ is almost surely generated under $Y_u = 1$ and we say that $x_u$ *indicates its state*. Consider $n$ to be the terminal time and note that any positive likelihood path, including $v(x_{1:n})$, the maximum likelihood one, must go through the state 1 at time $u$. This allows us to split the Viterbi alignment into $v^1$ and $v^2$, an alignment from time 1 through time $u$ and an alignment from time $u$ through time $n$, respectively. Namely, $v^1$ and $v^2$ maximize $\Lambda(y_{1:u}; x_{1:u})$ and $\Lambda(y_{u:n}; x_{u:n})$, the respective likelihoods. By concatenating $v^1$ with $v^2_{2:n-u+1}$ (removing the overlapping $v^2_1 = 1$), we obtain $v(x_{1:n})$ that maximizes $\Lambda(y_{1:n}; x_{1:n})$. Clearly, any additional observations $x_{n+1:m}$ do not change the fact that $x_u$ indicates its state. Hence, for any extension of $x_{1:n}$, the first part of the alignment is always $v^1$. Thus, any observation that indicates its state also fixes the beginning of the alignment. Since our HMM is a stationary process that has a positive probability of generating state-indicating observations, there will be infinitely many such observations almost surely. (The overlap $v_1^2 = 1$ is surely a nuisance since $v_{2:n-u+1}^2$ maximizes $\Lambda(y_{u+1:n}; x_{u+1:n})$ with the initial distribution $\pi$ replaced by $(p_{1j})_{j \in S}$.)

### 3.1. Nodes

The above example is rather exceptional and we next define nodes to generalize the idea of state-indicating observations.

First, consider the *scores*

$$\delta_u(l) \stackrel{\text{def}}{=} \max_{y_{1:u-1} \in S^{u-1}} \Lambda((y_{1:u-1}, l); x_{1:u}), \tag{3.1}$$

defined for all $u \geq 1$, $x_{1:u} \in \mathcal{X}^u$ and states $l$ in $S$. Thus, $\delta_u(l)$ is the maximum of the likelihood of the paths terminating at $u$ in state $l$. Note that $\delta_1(l) = \pi_l f_l(x_1)$. The recursion

$$\delta_{u+1}(j) = \max_{l \in S}(\delta_u(l) p_{lj}) f_j(x_{u+1}) \qquad \text{for all } u \geq 1 \text{ and } j \in S \tag{3.2}$$

helps to verify that $\mathcal{V}(x_{1:n})$, the set of all the Viterbi alignments, can be written as follows:

$$\mathcal{V}(x_{1:n}) = \{v \in S^n : \forall i \in S, \delta_n(v_n) \geq \delta_n(i) \text{ and } \forall u : 1 \leq u < n, v_u \in t(u, v_{u+1})\}, \tag{3.3}$$

where $t(u,j) \stackrel{\text{def}}{=} \{l \in S : \forall i \in S \ \delta_u(l) p_{lj} \geq \delta_u(i) p_{ij}\}$ for every $u = 1, \ldots, n$.

Thus, using (3.2), the Viterbi algorithm in its forward pass calculates $\delta_u(i)$, $i = 1, \ldots, K$, $u = 1, \ldots, n$, and stores maximizers $l \in t(u,j)$ (with some tie-breaking rule) to yield $\delta_{u+1}(j) = \delta_u(l) p_{lj} f_j(x_{u+1})$. The final alignment can then be found by backtracking as follows: $v_n \in \arg\max_{i \in S} \delta_n(i)$, $v_u \in t(u, v_{u+1})$, $u = n-1, \ldots, 1$.

**Definition 3.1.** *Given $x_{1:u}$, the first $u$ observations, the observation $x_u$ is said to be an $l$-node (of order 0) if*

$$\delta_u(l) p_{lj} \geq \delta_u(i) p_{ij} \qquad \text{for all } i, j \in S. \tag{3.4}$$

*We also say that $x_u$ is a node (of order 0) if it is an $l$-node for some $l \in S$. We say that $x_u$ is a strong node if the inequalities in (3.4) are strict for every $i, j \in S$, $i \neq l$. Definition 3.2 below generalizes this one by including nodes of positive orders.*

Clearly, if $x_u$ is an $l$-node, then $l \in t(u,j)$ for all $j \in S$ (see Figure 1). Consequently, if $x_{1:u}$ is such that $x_u$ is an $l$-node, then there exists $v(x_{1:n}) \in \mathcal{V}(x_{1:n})$ with $v(x_{1:n})_u = l$, which guarantees (the existence of) a fixed alignment up until $u$. If the node is strong, then all the Viterbi alignments must coalesce at $u$. Thus, the concept of strong nodes circumvents the inconveniences caused by the non-uniqueness. Namely, regardless of how the ties are broken, every alignment is forced into $l$ at $u$ and any tie-breaking rule would suffice for the purpose of obtaining the fixed alignments. However tempting, strong nodes, unlike the general ones, are quite restrictive. Indeed, suppose our model allows for $A$ with $P_1(A) > 0$ and $P_l(A) = 0$, for $l = 2, \ldots, K$. Hence, for almost every $x_u \in A$, we have $\delta_u(1) > 0$ and $\delta_u(i) = 0$ for every $i \in S$, $i \neq 1$. Thus, (3.4) holds and $x_u$ is a 1-node. If, in addition, $p_{1j} > 0$ for every $j \in S$, then for every $i, j \in S$, $i \neq 1$, the left-hand side of (3.4) is positive, whereas the right-hand side is 0, making $x_u$ a strong node. If, however, there is a $j$ such that $p_{1j} = 0$, which can easily happen if $K > 2$, then for this $j$, both sides are 0 and $x_u$ is no longer strong.

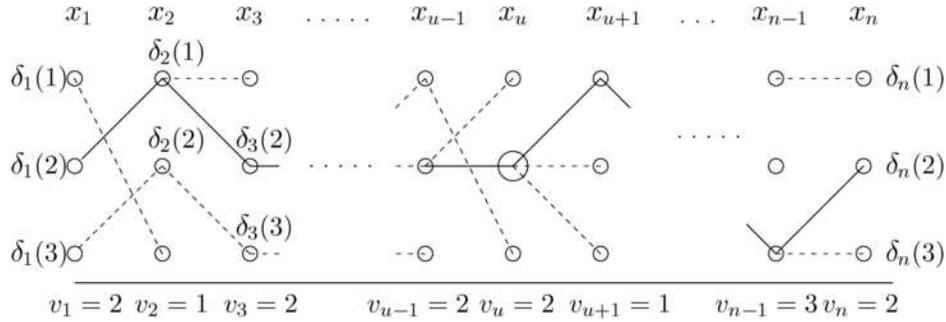

**Figure 1.** An example of the Viterbi algorithm in action. The solid line corresponds to the final alignment $v(x_{1:n})$. The dashed links are of the form $(k,l)-(k+1,j)$ with $l \in t(k,j)$ and are not part of the final alignment. For example, $(1,3)$–$(2,2)$ is because $3 \in t(1,2)$, $2 \in t(2,3)$. The observation $x_u$ is a 2-node since we have $2 \in t(u,j)$ for every $j \in S$. Also, note that $v(x_{1:u})$ is *fixed*, that is, $v(x_{1:u}) = v(x_{1:n})_{1:u}$.

The concept of nodes (including higher order nodes to be defined below) is essentially the same as 'crossing Viterbi paths' of [7] or 'meeting times/states' [5], where the existence of strong nodes is proved implicitly. The above works assume that the entries of $\mathbb{P}$, the transition matrix, are positive, which excludes our previous example of $x_u$ being a node and not a strong node. Using the concept of nodes, let us briefly analyze the results of these works. In [7], there are two main theorems. In terms of nodes, Theorem 1 of [7] states the following. *Let $j_0 \in S$ be a recurrent state. Let $i_0 \in S$ be such that for all $i, j, k \in S$, $i \neq i_0$,*

$$P_{j_0}(\{x \in \mathcal{X} : p_{ji_0} f_{i_0}(x) p_{i_0 k} > p_{ji} f_i(x) p_{ik}\}) > 0. \tag{3.5}$$

*Then, almost every realization of HMM has infinitely many nodes.* Up to notation, the condition (3.5) above is stated as it appears in [7]. However, this theorem is proved in [7] under the following stronger condition (3.6) (in [6], the authors of [7] have recently confirmed this to be a misprint):

$$P_{j_0}(\{x \in \mathcal{X} : p_{ji_0} f_{i_0}(x) p_{i_0 k} > p_{ji} f_i(x) p_{ik} \; \forall i, j, k \in S, i \neq i_0\}) > 0. \tag{3.6}$$

To see how significantly this alteration weakens the theorem, let $A \subset \mathcal{X}$ be the set as in (3.6) and let us first show that any $x_u \in A$ is a strong $i_0$-node. Indeed, fix $i \in S, i \neq i_0$. There then exists $j$ (depending on $i$) such that $\delta_u(i) = \delta_{u-1}(j) p_{ji} f_i(x_u)$. Next, for every $k$, $\delta_u(i) p_{ik} = \delta_{u-1}(j) p_{ji} f_i(x_u) p_{ik}$ and thus

$$\delta_u(i) p_{ik} < \delta_{u-1}(j) p_{ji_0} f_{i_0}(x_u) p_{i_o k} \leq \max_j \delta_{u-1}(j) p_{ji_0} f_{i_o}(x_u) p_{i_0 k} = \delta_u(i_0) p_{i_0 k}.$$

Thus, (3.6) implies that every observation from $A$ is a strong node. Since $j_0$ is recurrent and $A$ has a positive $P_{j_0}$-probability, clearly there are almost surely infinitely many such nodes. The existence of $A$ satisfying (3.6), however, appears to be more of an exception than a rule. Note that (3.6) does not hold if $\mathbb{P}$ contains a 0 in every row or in every column. Another important example of HMMs for which $A$ satisfying (3.6) need not exist is the HMM with additive white Gaussian noise (Example 1 of [5, 7]). In fact, it is stated in [7] that the assumption of their Theorem 1 is satisfied for this model independently of the transition matrix. In [6], the authors of [5, 7] have recently confirmed accidental omissions of the intended positivity condition, which, from the example below, can be seen to be crucial for Theorem 1 of [7], as well as Theorems 3 and 6 of [5]. Also, note that the following example does not require that $\mathbb{P}$ contain zeros in every row or column and is hence substantially different from the example given above. Thus, let $K = 3$ and let $p_{13} = 0$ be the only zero entry of $\mathbb{P}$. This already rules out (3.6) for $i_0 = 1$ and $i_0 = 3$. Following [7], in the additive white Gaussian noise model, the emission density $f_i$ is univariate normal with mean $i = 1, 2, 3$ and variance 1. Let $x$ be such that

$$p_{j2} f_2(x) p_{2k} > p_{ji} f_i(x) p_{ik} \qquad \forall i, j, k \in S, i \neq 2.$$

In particular, with $j = 2$, $p_{22} f_2(x) p_{23} > p_{23} f_3(x) p_{33}$ and $p_{22} f_2(x) p_{21} > p_{21} f_1(x) p_{11}$. Hence,

$$\frac{f_2(x)}{f_3(x)} > \frac{p_{33}}{p_{22}}, \qquad \frac{f_2(x)}{f_1(x)} > \frac{p_{11}}{p_{22}}. \tag{3.7}$$

Since $p_{11}$ and $p_{33}$ are both positive, one can easily find $p_{22} > 0$ sufficiently small for (3.7) to fail, implying that $i_0 \neq 2$. Therefore, (3.6), the (corrected) hypothesis of Theorem 1 of [7], which is also the hypothesis of Theorem 3 of [5], need not hold for the HMM with the additive Gaussian noise and $\mathbb{P}$ general.

We next extend the notion of nodes (Definition 3.1) to account for the fact that a general ergodic $\mathbb{P}$ can have a zero in every row, in which case nodes of order 0 need not exist. Indeed, suppose that $x_{1:u}$ is such that $\delta_u(i) > 0$ for every $i \in S$. In this case, (3.4) implies that $p_{lj} > 0$ for every $j \in S$ (the $l$th row of $\mathbb{P}$ must be positive) and (3.4) is equivalent to $\delta_u(l) \geq \max_i(\max_k(\frac{p_{ik}}{p_{lk}}) \delta_u(i))$.

First, we introduce $p_{ij}^{(r)}(u)$, the maximum likelihood of the paths connecting states $i$ and $j$ at times $u$ and $u + r$, respectively. Thus, for each $u \geq 1$ and $r \geq 1$, let

$$p_{ij}^{(r)}(u) \stackrel{\text{def}}{=} \max_{q_{1:r} \in S^r} p_{iq_1} f_{q_1}(x_{u+1}) p_{q_1 q_2} f_{q_2}(x_{u+2}) p_{q_2 q_3} \cdots p_{q_{r-1} q_r} f_{q_r}(x_{u+r}) p_{q_r j}.$$

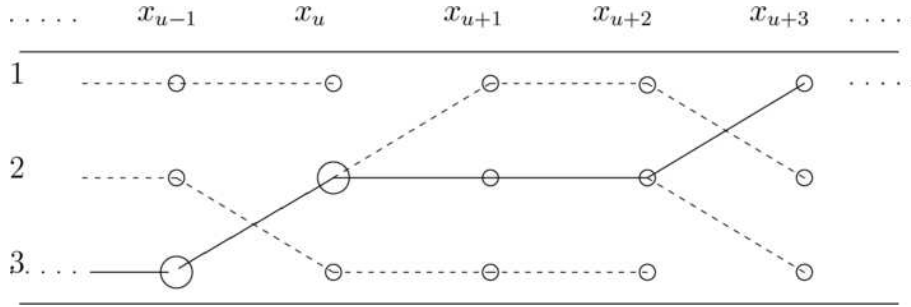

**Figure 2.** $x_u$ is a 2nd order 2-node, $x_{u-1}$ is a 3rd-order 3-node. Any alignment $v(x_{1:n})$ has $v(x_{1:n})_u = 2$.

Also, note that $p_{ij}^{(r)}(u) = \max_{q \in S} p_{iq}^{(r-1)}(u) f_q(x_{u+r}) p_{qj}$, where $p_{ij}^{(0)}(u) \stackrel{\text{def}}{=} p_{ij}$, $u \geq 1$. Recursion (3.2) then generalizes as follows: for all $r > u \geq 1$, for each $j \in S$, $\delta_{u+1}(j) = \max_{i \in S} (\delta_{u-r}(i) p_{ij}^{(r)}(u-r)) f_j(x_{u+1})$.

**Definition 3.2.** *Let $1 \leq r < n$, $1 \leq u \leq n - r$ and let $l \in S$. Given $x_{1:u+r}$, the first $u + r$ observations, $x_u$ is said to be an $l$-node of order $r$ if*

$$\delta_u(l) p_{lj}^{(r)}(u) \geq \delta_u(i) p_{ij}^{(r)}(u) \qquad \text{for all } i, j \in S. \tag{3.8}$$

*$x_u$ is said to be an $r$th-order node if it is an $r$th-order $l$-node for some $l \in S$. $x_u$ is said to be a* strong node of order $r$ *if the inequalities in* (3.8) *are strict for every $i, j \in S$, $i \neq l$.*

Note that any $r$th-order node is also a node of order $r'$ for any integer $r \leq r' < n$ and thus, by the order of a node, we will mean the minimal such $r$. Also, note that for $K = 2$, a node of any order is a node of order 0. Hence, positive order nodes only emerge for $K \geq 3$. If $x_u$ is an $l$-node of order $r$, then regardless of what the observations after $x_{u+r}$ are, $x_u$ remains an $l$-node of order $r$. Moreover, it follows from a decomposition of $\mathcal{V}(x_{1:n})$ similar to that of (3.3) that there exists $v(x_{1:n}) \in \mathcal{V}(x_{1:n})$ such that $v(x_{1:n})_u = l$. The difference between nodes (of order 0) and nodes of positive order $r$ is that for $v(x_{1:n})_u = l$ to hold, $u$ needs to be at least $r$ steps before $n$ $(n > u + r)$. Otherwise, for $m$ such that $u < m \leq u + r$, it might happen that no alignment $v(x_{1:m}) \in \mathcal{V}(x_{1:m})$ satisfies $v(x_{1:m})_u = l$. The role of higher order nodes is similar to that of nodes. Namely, provided a proper tie-breaking rule is given the existence of a higher order node $x_u$ ensures the existence of a fixed alignment up to $u$. At the same time, allowing nodes of higher orders removes the positivity restriction on rows of $\mathbb{P}$.

Although implicit (and defined relative to a fixed and global tie-breaking rule), nodes of orders possibly higher than 0 are also a main tool in [5, 7]. Specifically, statements $K'$ and $K''$, underpinning the main results of [7], are interpreted in terms of nodes as follows. $K'$: *almost every realization of an HMM has infinitely many (variable order) nodes.* (The node orders $r_1$, $r_2$, ... in $K'$ can depend on the realization $x_{1:\infty}$ and hence need not be almost surely bounded.) $K''$: *almost every realization of an HMM has infinitely many nodes of order* 0. (Thus, $K'$ implies $K''$ and for $K = 2$, $K'$ is equivalent to $K''$.) Lemmas 3.1 and 3.2 below give significantly stronger results, which also allow for an algorithmic construction of infinite piecewise alignments.

### 3.2. Piecewise alignment

Let $x_{1:n}$ be such that $x_{u_i}$ is an $l_i$-node of order $r$, $1 \leq i \leq k$, for some $k < n$ and assume that $u_k + r < n$ and $u_{i+1} > u_i + r$ for all $i = 1, 2, \ldots, k-1$. Such nodes are said to be *separated*. It follows from the definition of nodes that there exists a Viterbi alignment $v_{1:n} \in \mathcal{V}(x_{1:n})$ such that $v_{u_i} = l_i$ for every $i = 1 \leq k$. Indeed, Definition 3.2 immediately implies the existence of a Viterbi alignment $v'_{1:n} \in \mathcal{V}(x_{1:n})$ with $v'_{u_k} = l_k$. The same definition and optimality of backtracking by the Viterbi algorithm imply that $(w_{1:u_{k-1}+r}, v'_{u_{k-1}+r+1:n}) \in \mathcal{V}(x_{1:n})$ for some prefix $w_{1:u_{k-1}+r}$ with $w_{u_{k-1}} = l_{k-1}$. Continuing in this manner down to node $x_{u_1}$, we exhibit $v_{1:n}$ with $v_{u_i} = l_i$, $i = 1, 2, \ldots, k$.

Let us discuss the assumption $u_{i+1} > u_i + r$, $i = 1, 2, \ldots, k-1$. The fact that $x_{u_i}$ is an $r$th-order $l_i$-node guarantees that when backtracking from $u_i + r$ down to $u_i$, ties can be broken in such a way that, regardless of the values of $x_{u+r+1:n}$ and how ties are broken in between $n$ and $u_i + r$, the alignment goes through $l_i$ at $u_i$. At the same time, segment $u_i, \ldots, u_i + r$ is 'delicate', that is, unless $x_{u_i}$ is a strong node, breaking ties arbitrarily on $u_i, \ldots, u_i + r$ can result in $v(x_{1:n})_{u_i} \neq l_i$. Hence, when neither $x_{u_i}$ nor $x_{u_{i+1}}$ is strong and $u_{i+1} \leq u_i + r$, breaking ties in favor of $x_{u_i}$ can result in $v_{u_{i+1}} \neq l_{i+1}$. Note that such a pathological situation is impossible if $r = 0$ and might be rare in practice for $r > 0$. Finally, note that this assumption is not restrictive since it is always possible to choose from any sequence of nodes a subsequence of nodes that are separated.

To formalize the piecewise construction introduced above, let

$$\mathcal{W}^l(x_{1:n}) = \{v \in S^n : v_n = l, \Lambda(v; x_{1:n}) \geq \Lambda(w; x_{1:n}) \ \forall w \in S^n : w_n = l\},$$
$$\mathcal{V}^l(x_{1:n}) = \{v \in \mathcal{V}(x_{1:n}) : v_n = l\}, \qquad \text{for all } n \geq 1, l \in S \text{ and } x_{1:n} \in \mathcal{X}^n,$$

be the sets of maximizers of the constrained likelihood and the subset of maximizers of the (unconstrained) likelihood, respectively, all elements of which go through $l$ at $u$. Note that, unlike $\mathcal{W}^l(x_{1:n})$, $\mathcal{V}^l(x_{1:n})$ might be empty. It can be shown that $\mathcal{V}^l(x_{1:n}) \neq \varnothing$ implies that $\mathcal{V}^l(x_{1:n}) = \mathcal{W}^l(x_{1:n})$. Also, let the subscript $(l)$ stand for using $(p_{li})_{i \in S}$ as the initial distribution in place of $\pi$. Thus, the sets $\mathcal{V}_{(l)}(x_{1:n})$ and $\mathcal{W}^m_{(l)}(x_{1:n})$, $m \in S$, will also be used.

The piecewise construction can be formulated as follows. Suppose that there exist $l_1, \ldots, l_k \in S$ and $u_1, \ldots, u_k \geq 1$, $r_1, \ldots, r_k \geq 0$ with $u_1 + r_1 < u_2 + r_2 < \cdots < u_k + r_k < n$ such that $x_{u_i}$ is an $l_i$-node of order $r_i$ for every $i \leq k$. There then exists an alignment $v(x_{1:n}) = (v^1, \ldots, v^{k+1}) \in \mathcal{V}(x_{1:n})$, where $v^1 \in \mathcal{W}^{l_1}(x_{1:u_1})$,

$$v^i \in \mathcal{W}^{l_i}_{(l_{i-1})}(x_{u_{i-1}+1:u_i}), \qquad 2 \leq i \leq k, \quad \text{and} \quad v^{k+1} \in \mathcal{V}_{(l_k)}(x_{u_k+1:n}). \tag{3.9}$$

Moreover, for every $i = 1, 2, \ldots, k$, $w(i) \stackrel{\text{def}}{=} (v^1, \ldots, v^i) \in \mathcal{V}^{l_i}(x_{1:u_i})$. Thus, when a node is observed at time $u_k$, the alignment up to $u_k$ becomes fixed, yielding natural extensions of finite alignments for $n \to \infty$. Besides providing the tool for the asymptotic analysis, the piecewise construction is also of computational significance. Indeed, note that once $x_{u_1}$ has been recognized to be a node and $w(1)$ has been constructed, the memory allocated for storing $x_{1:u_1}$ and $t(u, j)$ (see (3.3)) for $u \leq u_1$ and $j \in S$ is no longer needed and can be freed.

Thus, if $x_{1:\infty}$ has infinitely many nodes $\{x_{u_k}\}_{k \geq 1}$ that are separated, then $v(x_{1:\infty})$, an *infinite piecewise alignment based on the node times* $\{u_k(x_{1:\infty})\}_{k \geq 1}$ can be defined as follows. If the sets $\mathcal{W}^{l_i}_{(l_{i-1})}(x_{u_{i-1}+1:u_i})$, $i \geq 2$, as well as $\mathcal{W}^{l_1}(x_{1:u_1})$ are singletons, then (3.9) immediately defines a unique infinite alignment $v(x_{1:\infty}) = (v^1(x_{1:u_1}), v^2(x_{u_1+1:u_2}), \ldots)$. Otherwise, ties must be broken. In order for our infinite alignment process to be regenerative, a natural consistency condition must be imposed on rules to select unique $v(x_{1:n})$ from $\mathcal{W}^{l_1}(x_{1:u_1}) \times \mathcal{W}^{l_2}_{(l_1)}(x_{u_1+1:u_2}) \times \cdots \times \mathcal{W}^{l_k}_{(l_{k-1})}(x_{u_{k-1}+1:u_k}) \times \mathcal{V}_{(l_k)}(x_{u_k+1:n})$. Resulting infinite alignments, as well as decoding $v: \mathcal{X}^\infty \to S^\infty$ based on such alignments, will be called *proper*. This condition is, perhaps, best understood by the following example. Suppose, for some $x_{1:5} \in \mathcal{X}^5$, that $\mathcal{W}^1_{(1)}(x_{1:5}) = \{12211, 11211\}$ and suppose that the tie is broken in favor of 11211. Now, whenever $\mathcal{W}^1_{(l)}(x'_{1:4})$ contains $\{1221, 1121\}$, we naturally require that 1221 not be selected. In particular, we break the tie in $\mathcal{W}^1_{(1)}(x_{1:4}) = \{1221, 1121\}$ by selecting 1121. Subsequently, 112 is selected from $\mathcal{W}^2_{(1)}(x_{1:3}) = \{122, 112\}$, and so on. It can be shown that a decoding by piecewise alignment (3.9) with ties broken in favor of min (or max) under the reverse lexicographic ordering of $S^n$, $n \in \mathbb{N}$, is a proper decoding.

**Example 2 (Mixtures revisited).** Consider the mixture model as in Example 1. In this case, an observation $x_u$ is an $l$-node if and only if $\delta_u(l) \geq \delta_u(i)$ for every $i \in S$. In particular, this implies that every observation is an $l$-node (of order 0) for some $l \in S$. Recursion (3.2) can then be written for any $u \geq 2$ and $i \in S$ as $\delta_u(i) = \max_{j \in S} \delta_{u-1}(j) \pi_i f_i(x_u) = c \pi_i f_i(x_u)$, where $c$ does not depend on $i$. Hence, $x_u$ is an $l$-node if and only if $\pi_l f_l(x_u) \geq \pi_i f_i(x_u)$ for all $i \in S$. Therefore, the alignment can be obtained component-wise: $v(x_{1:n}) = (v(x_1), \ldots, v(x_n))$, where

$$v(x) = \arg\max_{i \in S} \pi_i f_i(x). \tag{3.10}$$

Clearly, the alignment is proper if the ties in (3.10) are broken consistently, that is, if $v(x)$ is indeed a well-defined function of $x$.

Example 2 helps to understand the necessity of breaking ties consistently. If our sole goal were to construct infinite alignments, then any piecewise (not necessarily proper) alignment would suffice. However, the existence of $Q_l(\psi)$, $l \in S$, requires more. Indeed, suppose that the right-hand side of (3.10) is not unique for some $x$, an atom of, say $\hat{P}^n_1$, as defined in (2.3). If the selection in (3.10) is consistent, say, $v(x) = 1$, then, in the limit, $x$ will also be an atom of $Q_1(\psi)$. Otherwise, if ties in (3.10) are broken arbitrarily, then the limiting measures might not exist at all.

Also, note that we break ties locally, that is, within individual intervals $u_{i-1}+1,\ldots,u_i$, $i\geq 2$, enclosed by the adjacent nodes. This is in contrast to global ordering of $\mathcal{V}(x_{1:\infty})$, such as the one in [5, 7], which ignores decomposition (3.9). A global rule can fail to produce an infinite alignment going through infinitely many nodes unless the nodes are strong (as assumed in [5, 7]).

### 3.3. Barriers

To test whether $x_u$ is a node of order $r$ requires the entire realization $x_{1:u+r}$ (Definition 3.2). In particular, for an arbitrary prefix $x'_{1:w} \in \mathcal{X}^w$ and $m < u$, the $(w+m+1)$th element of $(x'_{1:w}, x_{u-m:u+r})$ need not be a node relative to $(x'_{1\ldots w}, x_{u-m:u+r})$, even when $x_u$ is a node of order $r$ relative to $x_{1:u+r}$. We show below that typically, a block $x^b_{1:k} \in \mathcal{X}^k$ ($k \geq r$) can be found such that for any $w \geq 1$ and any $x'_{1:w} \in \mathcal{X}^w$, the $(w+k-r)$th element of $(x'_{1:w}, x^b_{1:k})$ is a node of order $r$ (relative to $(x'_{1:w}, x^b_{1:k})$). Sequences $x^b_{1:k}$ that ensure the existence of such persistent nodes will be called *barriers*.

**Definition 3.3.** *Given $l \in S$, $x^b_{1:k} \in \mathcal{X}^k$ is called a (strong) $l$-barrier of order $r \geq 0$ and length $k \geq 1$ if, for any $w \geq 1$ and every $x'_{1:w} \in \mathcal{X}^w$, $(x'_{1:w}, x^b_{1:k})$ is such that $(x'_{1:w}, x^b_{1:k})_{w+k-r}$ is a (strong) $l$-node of order $r$.*

Note that any observation from the set $A$ considered in (3.6) is a barrier of length 1. In particular, any observation that indicates a state is a barrier of length 1.

Next, we state and discuss Lemmas 3.1 and 3.2, the first of the two main results of this paper. First, let $G_l = \bigcap_{G\text{-closed}, P_l(G;\theta_l)=1} G$ denote the support of the family $P_l(\theta_l)$, $\theta_l \in \Theta_l$, for all $l \in S$.

**Definition 3.4.** *We call a subset $C \subset S$ a cluster, if the following conditions are satisfied:*

$$\min_{j \in C} P_j\left(\bigcap_{i \in C}(G_i \cap \{x \in \mathcal{X} : f_i(x) > 0\})\right) > 0 \quad and \quad P_j\left(\bigcap_{i \in C} G_i\right) = 0 \quad \forall j \notin C.$$

Hence, a cluster is a maximal subset of states such that $G_C = \bigcap_{i \in C} G_i$ is 'detectable'. Distinct clusters need not be disjoint and a cluster can consist of a single state. In this latter case, such a state is not hidden since it is indicated by any observation which it emits. When $K = 2$, $S$ is the only cluster possible since otherwise, all observations would reveal their states and the underlying Markov chain would cease to be hidden. In practice, many other HMMs have the entirity of $S$ as their (necessarily unique) cluster.

The proof of the following lemma is rather technical and can be found in [26], Appendix 5.1, pages 26–39.

**Lemma 3.1.** *Assume that for each state $l \in S$,*

$$P_l\left(\left\{x \in \mathcal{X} : f_l(x)\max_{j \in S}(p_{jl}) > \max_{i \in S, i \neq l}\left(f_i(x)\max_{j \in S}(p_{ji})\right)\right\}\right) > 0. \tag{3.11}$$

*Moreover, assume that there exists a cluster $C \subset S$ and an integer $m < \infty$ such that the mth power of the substochastic matrix $Q = (p_{ij})_{i,j \in C}$ is strictly positive. Then, for some integers $M$ and $r$, $M > r \geq 0$, there exist $B = B_1 \times \cdots \times B_M \subset \mathcal{X}^M$, $q_{1:M} \in S^M$ and $l \in S$ such that every $x_{1:M}^b \in B$ is an l-barrier of order $r$ (and length $M$), $q_{M-r} = l$, $\mathbf{P}(X_{1:M} \in B | Y_{1:M} = q_{1:M}) > 0$ and $\mathbf{P}(Y_{1:M} = q_{1:M}) > 0$.*

Lemma 3.1 implies that $\mathbf{P}(X_{1:M} \in B) > 0$. Also, since every element of $B$ is a barrier of order $r$, the ergodicity of $X$ therefore guarantees that almost every realization of $X_{1:\infty}$ contains infinitely many $l$-barriers of order $r$. Hence, almost every realization of $X_{1:\infty}$ also has infinitely many $l$-nodes of order $r$.

Let us briefly analyze (3.11) and the existence of a cluster $C$ assumed in Lemma (3.1). First, consider the case when $S$ itself is a cluster. This occurs, for example, if the supports of all the emission distributions coincide. Then, the substochastic matrix $(p_{ij})_{i,j \in C} = \mathbb{P}$ and aperiodicity of $\mathbb{P}$ implies that $\mathbb{P}^m$ is strictly positive for some power $m$. Hence, the cluster assumption is satisfied in this case. Our cluster assumption essentially generalizes assumption A1 of [5, 7], which requires $\mathbb{P}$, the transition matrix, to be strictly positive and the supports $G_i$ to be all equal. As already pointed out, the assumption of strict positivity of $\mathbb{P}$ becomes rather restrictive when $K > 2$. Moreover, [26], Example 3.11, shows that the cluster assumption is not only sufficient but also *necessary* for nodes (and barriers) to exist. We also point out that the proof of the existence of nodes in [5] (Theorem 2) heavily relies on the supports being equal, which is also crucial for assumption A2 [5, 7] and which is not assumed in Lemma 3.1.

Note that (3.11) basically says that for every state $l \in S$, there is a set where the measure $P_l(\theta_l)$ 'dominates', that is, $\{x \in \mathcal{X} : f_l(x)\max_{j \in S} p_{jl} > \max_{i \in S, i \neq l}(f_i(x)\max_{j \in S} p_{ji})\}$ is of positive $\lambda$-measure. We are not aware of any HMMs used in practice for which this assumption does not hold. Moreover, for many models (see Example 3 below), it is actually sufficient for proving the existence of barriers that (3.11) holds for at least one state $l$, which, provided that the emission distributions $P_l(\theta_l)$, $l \in S$, are all distinct, is always the case. Also, note that for the mixture model, (3.11) simplifies to $P_l(\{x : f_l(x)\pi_l > f_i(x)\pi_i, \forall i \neq l\}) > 0$ and that assumption (3.11) is weaker than (3.6) since the latter implies that

$$P_{i_0}\left(\left\{x \in \mathcal{X} : f_{i_0}(x)\max_{j \in S} p_{ji_0} > \max_{i \in S, i \neq i_0}\left(f_i(x)\max_{j \in S} p_{ji}\right)\right\}\right) > 0.$$

**Example 3 ($K = 2$).** $S = \{1, 2\}$ is the only cluster. Assume $\mathbb{P}$ to be strictly positive. Thus, the cluster assumption of Lemma 3.1 is fulfilled. Assume $P_1(\theta_1)$ and $P_2(\theta_2)$ to be distinct. Following [5], consider the following three cases. Case 1: $p_{11} > p_{21}$ (equivalently, $p_{22} > p_{12}$); case 2: $p_{11} < p_{21}$ (equivalently, $p_{22} < p_{12}$); case 3: $p_{11} = p_{21}$ (equivalently, $p_{22} = p_{12}$). Note that since $\lambda(\{x \in \mathcal{X} : f_1(x) \neq f_2(x)\}) > 0$ (the two emission distributions differ), the sets $\mathcal{X}_1 \stackrel{\text{def}}{=} \{x \in \mathcal{X} : f_1(x)p_{11} > f_2(x)p_{22}\}$, $\mathcal{X}_2 \stackrel{\text{def}}{=} \{x \in \mathcal{X} : f_1(x)p_{11} < f_2(x)p_{22}\}$ satisfy

$$\lambda(\mathcal{X}_1) > 0 \text{ or } \lambda(\mathcal{X}_2) > 0. \tag{3.12}$$

Without loss of generality, assume $p_{11} \geq p_{22}$, hence $\lambda(\mathcal{X}_1) > 0$. It is then not hard to exhibit strong 1-barriers in case 1. Indeed, in this case, a Viterbi path $v(x_{1:n})$ can switch states only at nodes, that is, $v(x_{1:n})_{u:u+1} = (l, j)$, $l \neq j$, implies that $x_u$ is a strong $l$-node. An integer $k$ can then be chosen sufficiently large for any sequence $z_{1:k} \in \mathcal{X}_1^k$ to be a strong 1-barrier. Suppose that this were not the case and hence that no $z_i$, $1 \leq i \leq k$, would be a 1-node. It could then be shown that no $z_i$ could be a 2-node either, hence corresponding $k$-segments of Viterbi paths $v(x_{1:n})$, $n > k$, would have to be constant, namely all 1's or all 2's. However, $k$ is so large that segment $211\ldots12$ is more optimal than $22\ldots2$, implying the presence of a strong 1-node.

Thus, in case 1, the occurrence of infinitely many barriers (or nodes) does not require any additional assumptions. In particular, assumptions A1 (the supports being equal) and A2 (log-ratio of the densities being square-integrable) of [5, 7] are unnecessary for proving the results of Theorems 7, 8 and 9 of [5]. Furthermore, assumption (3.11) of Lemma 3.1 is, in this case, equivalent to the conjunction of $\lambda(\mathcal{X}_1) > 0$ and $\lambda(\mathcal{X}_2) > 0$. Thus, Lemma 3.1 can be further strengthened in this case to guarantee that almost every realization of the HMM has infinitely many both 1- and 2-barriers. Alternatively, assumption (3.11) can be relaxed to (3.12) in this case, as well as in many other practical situations, for Lemma 3.1 to still guarantee at least one type of barrier.

Next, consider case 2. Lemma 3.1 says that when both sets

$$\begin{aligned}
\mathcal{X}_1 &\stackrel{\text{def}}{=} \{x \in \mathcal{X} : f_1(x)p_{21} > f_2(x)p_{12}\}, \\
\mathcal{X}_2 &\stackrel{\text{def}}{=} \{x \in \mathcal{X} : f_1(x)p_{21} < f_2(x)p_{12}\}
\end{aligned} \tag{3.13}$$

have positive $\lambda$-measure, then almost every realization $x_{1:\infty}$ includes infinitely many barriers. One can show that these barriers are the elements of the set $B = \mathcal{X}_1 \times \mathcal{X}_2 \times \mathcal{X}_1 \times \cdots \times \mathcal{X}_2$. Indeed, it can be shown that the absence of nodes in a generic subsequence $x_{t:t+T}$ would imply optimality of the likelihood motif $p_{ba}f_a(x_t)p_{ab}f_b(x_{t+1})$, $a, b \in S$, $a \neq b$. However, if $x_{t:t+T} \in \mathcal{X}_b \times \mathcal{X}_a \times \mathcal{X}_b \times \cdots$ and $T$ is sufficiently large, then this motif will no longer be optimal, hence a node inside $x_{t:t+T}$. In [28], we additionally show that barriers (or nodes) also exist in case 2, even if only one of the sets in (3.13) has positive measure. Since a typical Viterbi path in case 2 oscillates between the states (as also acknowledged in [5]), case 2 is not similar to case 1, requiring a different approach to prove the existence of barriers (or nodes) under the weakened assumption $\max\{\lambda(\mathcal{X}_1), \lambda(\mathcal{X}_2)\} > 0$. This also explains why we generally ($K \geq 2$) require (3.11) to hold for more than one state. In [5], the author reports similar results, Theorems 10 and 11, without proofs, alleging that the omitted proofs are "very similar" to the respective proofs of Theorems 7 and 8 of [5]. We are convinced that proving Theorem 10 of [5] requires an approach different from that of the proof of Theorem 7 in [5].

Finally, case 3 is the mixture model with weights $\pi_1 = p_{11} = p_{21}$, $\pi_2 = p_{22} = p_{12}$. Every observation is now a node (Example 2). Again, if $\lambda(\{f_1 \neq f_2\}) > 0$ holds, then so does (3.12), say, with the first of its statements. Every element of $\{x \in \mathcal{X} : f_1(x)\pi_1 > f_2(x)\pi_2\}$ is then a strong 1-barrier of order 0 and length 1. Therefore, unlike in Theorems 12, 13 and 14 of [5], the existence of infinitely many barriers (nodes) again follows with no additional assumptions.

In summary, barriers allow us to prove, relatively easily, the existence of infinitely many nodes. Although the existence of barriers is rather obvious for $K = 2$, the CLT-based proof of [7], Theorem 2, does not apply if $K > 2$, necessitating generalizations such as Lemma 3.1.

For certain technical reasons, instead of extracting subsequences of separated nodes from general infinite sequences of nodes guaranteed by Lemma 3.1, we achieve node separation by adjusting the notion of barriers. Namely, note that two $r$th order $l$-barriers $x_{j:j+M-1}$ and $x_{i:i+M-1}$ might be in $B$ with $j < i \leq j+r$, implying that the associated nodes $x_{j+M-r-1}$ and $x_{i+M-r-1}$ are not separated. Thus, we impose on $B$ the following condition:

$$x_{j:j+M-1}, x_{i:i+M-1} \in B, \quad i \neq j \implies |i-j| > r. \tag{3.14}$$

If (3.14) holds, then we say that the barriers from $B \subset \mathcal{X}^M$ are *separated*. This is often easy to achieve by a simple extension of $B$, as shown in the following example. Suppose that there exists $x \in \mathcal{X}$ such that $x \notin B_m$ for all $m = 1, 2, \ldots, M$. All elements of $B^* \stackrel{\text{def}}{=} \{x\} \times B$ are evidently barriers and, moreover, they are now separated. The following lemma incorporates a more general version of the above example (see [26], Appendix 5.2, pages 39–40, for proof).

**Lemma 3.2.** *Suppose that the assumptions of Lemma 3.1 are satisfied. Then, for some integers $M$ and $r$, $M > r \geq 0$, there exist $B = B_1 \times \cdots \times B_M \subset \mathcal{X}^M$, $q_{1:M} \in S^M$ and $l \in S$ such that every $x_{1:M}^b \in B$ is a separated $l$-barrier of order $r$ (and length $M$), $q_{M-r} = l$, $\mathbf{P}(X_{1:M} \in B | Y_{1:M} = q_{1:M}) > 0$ and $\mathbf{P}(Y_{1:M} = q_{1:M}) > 0$.*

## 4. The alignment process

For the rest of this work, we adopt the assumptions of Lemma 3.2 to guarantee that almost every realization of HMM has infinitely many separated barriers. Every such barrier contains a node. Note that both the barrier and the node encapsulated in it are therefore observable via testing the running $M$-tuples of $X_{1:\infty}$ for membership in $B$. Based on such nodes, we define $v : \mathcal{X}^\infty \to S^\infty$ to be a proper decoding by piecewise alignment (3.9) (and $v(x_{1:\infty})_i = 1$, $i \geq 1$, for $x_{1:\infty}$ that do not have infinitely many $B$-barriers). Next, we study properties of the random *alignment* process $V_{1:\infty} \stackrel{\text{def}}{=} v(X_{1:\infty})$.

Let $M \geq 0$, $B \subset \mathcal{X}^M$, $r \geq 0$, $l \in S$ and $q = q_{1:M} \in S^M$, as promised by Lemma 3.2. For every $n \geq 1$, $\mathbf{P}(Y_{n:n+M-1} = q) > 0, \gamma^* \stackrel{\text{def}}{=} \mathbf{P}(X_{n:n+M-1} \in B | Y_{n:n+M-1} = q) > 0$, hence every $x_{n:n+M-1} \in B$ is a separated $l$-barrier of order $r$. Next, define, for all $n \geq 1$,

$$U_n \stackrel{\text{def}}{=} X_{n:n+M-1}, \qquad D_n \stackrel{\text{def}}{=} Y_{n:n+M-1}, \qquad \mathcal{F}_n \stackrel{\text{def}}{=} \sigma(Y_{1:n}, X_{1:n}), \text{ as well as}$$

stopping times $\nu_0, \nu_1, \nu_2, \ldots, \vartheta_0, \vartheta_1, \vartheta_2, \ldots$ of the filtration $\{\mathcal{F}_{n+M-1}\}_{n \geq 1}$:

$$\nu_0 \stackrel{\text{def}}{=} \min\{n \geq 1 : U_n \in B, D_n = q\},$$
$$\nu_i \stackrel{\text{def}}{=} \min\{n > \nu_{i-1} : U_n \in B, D_n = q\} \qquad \forall i \geq 1,$$
$$\vartheta_0 \stackrel{\text{def}}{=} \min\{n \geq 1 : U_n \in B\},$$
$$\vartheta_i \stackrel{\text{def}}{=} \min\{n > \vartheta_{i-1} : U_n \in B\} \qquad \forall i \geq 1,$$

with the convention that $\min \varnothing = 0$ and $\max \varnothing = -1$. Note that $\vartheta_i \leq \nu_i$, $i \geq 0$. Stopping times $\vartheta_i$ ($i \geq 0$) are observable via the $X$ process alone, whereas stopping times $\nu_i$ ($i \geq 0$) already require knowledge of the full process $(X_{1:\infty}, Y_{1:\infty})$. Also, note that $\nu_0$, $(\nu_{i+1} - \nu_i)$, $i \geq 0$, are independent and $(\nu_{i+1} - \nu_i)$, $i \geq 0$, are identically distributed. To every $\nu_i$, there corresponds an $l$-barrier of order $r$. This barrier extends over the interval $[\nu_i, \nu_i + M - 1]$ and $X_{\tau_i}$ is an $l$-node of order $r$, where $\tau_i \stackrel{\text{def}}{=} \nu_i + (M-1) - r$ for every $i \geq 0$. Define $T_0 \stackrel{\text{def}}{=} \tau_0$ and $T_i \stackrel{\text{def}}{=} \tau_i - \tau_{i-1} = \nu_i - \nu_{i-1}$ for every $i \geq 1$.

**Proposition 4.1.** $E(T_0) < \infty$ and $E(T_1) < \infty$.

**Proof.** We need to show that $E\nu_0 < \infty$ and $E(\nu_1 - \nu_0) < \infty$. Let us introduce the following non-overlapping block-valued processes $U_m^b$ and $D_m^b$, defined by $U_m^b = X_{(m-1)M+1:mM}$, $D_m^b = Y_{(m-1)M+1:mM}$, for all $m \geq 1$, and stopping times defined, for every $i \geq 1$, by

$$\nu_0^b \stackrel{\text{def}}{=} \min\{m \geq 1 : U_m^b \in B, D_m^b = q\},$$
$$\nu_i^b \stackrel{\text{def}}{=} \min\{m > \nu_{i-1}^b : U_m^b \in B, D_m^b = q\},$$
(4.1)

$$R_0^b \stackrel{\text{def}}{=} \min\{m > 1 : D_m^b = q\},$$
$$R_i^b \stackrel{\text{def}}{=} \min\{m > R_{i-1}^b : D_m^b = q\}.$$
(4.2)

The process $D^b$ is clearly a time-homogeneous, finite-state Markov chain. Since $Y_{1:\infty}$ is aperiodic and irreducible, so is $D^b$. Hence, $(D^b, U^b)$ is also an HMM.

Since $Y_{1:\infty}$ is also stationary (under $\pi$), $q$ occurs in every interval of length $M$ with the same positive probability (Lemma 3.2). In particular, $q$ belongs to the state space of $D^b$. Since $D^b$ is irreducible and its state space is finite, all of its states, including $q$, are positive recurrent. Hence, $E(R_0^b) < \infty$ and $E(R_1^b - R_0^b) < \infty$. The following bound ultimately yields the second statement: $E(\nu_1 - \nu_0) \leq E(\nu_1^b - \nu_0^b) = \frac{1}{\gamma^*} E(R_1^b - R_0^b) < \infty$. This bound is obtained by twice applying Wald's equation [3].

It can similarly be verified that $E(\nu_0^b) = \gamma^* E(R_0^b) + \frac{1-\gamma^*}{\gamma^*} E(R_1^b - R_0^b)$, which is again finite. Finally, $E\nu_0 \leq M(E\nu_0^b - 1) + 1 < \infty$. □

According to Proposition 4.1 above, $ET_i < \infty$ for every $i \geq 0$, implying that the random variables $T_0, T_1, \ldots$ form a delayed renewal process (for a general reference, see, e.g., [3]). In [5], the process $\tau$ and the expectation $ET_1$ are denoted by $S$ and $E(S_1|S_0)$, respectively. As the proof of Proposition 4.1 above shows, using the barriers, it is relatively easy to prove that $ET_1 < \infty$. On the other hand, without such a unifying concept, [5] must prove $E(S_1|S_0) < \infty$ separately for every case considered therein.

Next, let $u_0, u_1, \ldots$ be the locations of $r$th order $l$-nodes corresponding to the stopping times $\vartheta_i$, that is, $u_i = \vartheta_i + (M-1) - r$ for every $i \geq 0$. Clearly, for every $i \geq 0$, $\tau_i = u_j$ for some $j \geq i$. Also, since the barriers are separated, so are $(u_i)_{i \geq 0}$. Using these nodes, we build the alignment $v$ and thus extend the definitions of the empirical measures $\hat{P}_l^n(\psi, X_{1:n})$ given in (2.3) and the estimators of transition probabilities $\hat{p}_{ij}^n$ given in (2.2) for the general case of non-unique alignments. Specifically, given $X_{1:n}$, define $V'_{1:n} = v(X_{1:n})$ to be the (finite) piecewise proper alignment based on the $u_i$'s (and a consistent selection scheme) in accordance with (3.9). For each state $l \in S$ that appears in $V'_{1:n}$, define

$$\hat{P}_l^n(A; \psi, X_{1:n}) \stackrel{\text{def}}{=} \frac{\sum_{i=1}^n I_{A \times \{l\}}(X_i, V'_i)}{\sum_{i=1}^n I_{\{l\}}(V'_i)}, \qquad A \in \mathcal{B}.$$

For other $l \in S$ (i.e., $\sum_{i=1}^n I_{\{l\}}(V'_i) = 0$), define $\hat{P}_l^n(\psi, X_{1:n})$ to be an arbitrary probability measure.

Similarly, for every pair of states $l, j \in S$, we define

$$\hat{p}_{lj}^n(\psi, X_{1:n}) \stackrel{\text{def}}{=} \frac{\sum_{i=1}^{n-1} I_{\{l\}}(V'_i) I_{\{j\}}(V'_{i+1})}{\sum_{i=1}^{n-1} I_{\{l\}}(V'_i)}.$$

Again, if $\sum_{i=1}^{n-1} I_{\{l\}}(V'_i) = 0$, define $\hat{p}_{l \cdot}^n(\psi, X_{1:n})$ to be an arbitrary probability vector on $S$.

We shall next consider the 2-dimensional process $Z \stackrel{\text{def}}{=} (X_{1:\infty}, V_{1:\infty})$. Based on $Z$, for every $l \in S$, we also define auxiliary empirical measures $\hat{Q}_l^n$ and $(\hat{q}_{lj}^n)_{j \in S}$ as follows:

$$\hat{Q}_l^n(A, Z_{1:n}) \stackrel{\text{def}}{=} \frac{\sum_{i=1}^n I_{A \times \{l\}}(X_i, V_i)}{\sum_{i=1}^n I_{\{l\}}(V_i)} = \frac{\sum_{i=1}^n I_{A \times \{l\}}(Z_i)}{\sum_{i=1}^n I_{\{l\}}(V_i)}, \qquad A \in \mathcal{B},$$

$$\hat{q}_{lj}^n(Z_{1:n}) \stackrel{\text{def}}{=} \frac{\sum_{i=1}^{n-1} I_{\{l\}}(V_i) I_{\{j\}}(V_{i+1})}{\sum_{i=1}^{n-1} I_{\{l\}}(V_i)} \qquad \text{for every } j \in S.$$

As in the definition of $\hat{P}_l^n(\psi, X_{1:n})$, if $l \neq V_i$, $i = 1, \ldots, n$ ($i = 1, \ldots, n-1$), then $\hat{Q}_l^n(Z_{1:n})$'s ($\hat{q}_{l \cdot}^n(Z_{1:n})$'s) are defined arbitrarily. Note that, in general, $v(x_{1:\infty})_{1:n} \neq v(x_{1:n})$. However, the two are equal up to the last node occurring prior to $n$ and used in the construction of $v$. Thus, after that last node, $V'_i$ need no longer agree with $V_i$.

To prove the existence of $Q_l$ such that $\hat{P}_l^n(\psi, X_{1:n}) \Rightarrow Q_l(\psi)$ a.s., we first note that $Z$ is a regenerative process [3] with respect to the renewal times $(\tau_i)_{i \geq 0}$. This implies that $\hat{Q}_l^n(Z_{1:n}) \Rightarrow Q_l(\psi)$, a.s. Finally, since the difference between $\hat{Q}_l^n(Z_{1:n})$ and $\hat{P}_l^n(\psi, X_{1:n})$ vanishes as $n \to \infty$, we have $\hat{P}_l^n(\psi, X_{1:n}) \Rightarrow Q_l(\psi)$ almost surely. Similarly, we prove the almost sure convergence $\hat{p}_{lj}^n(\psi, X_{1:n}) \to q_{lj}(\psi)$.

The fact that the process $Z$ is regenerative is crucial and is the main result in [5], Theorem 2. That $X$ is regenerative immediately follows from the fact that for every $i \geq 0$, $Y_{\tau_i} = l$ and the $T_i$'s are renewal times. $V$ is regenerative because all the nodes occurring at $\tau_i$'s are used in the construction of $V_{1:\infty}$ via (3.9) and because decoding $V_{1:\infty}$ is proper. That is, for every $i \geq 1$, $V_{\tau_{i-1}+1:\tau_i} = v^j \in \mathcal{W}_{(l)}^l(X_{\tau_{i-1}+1:\tau_i})$ for some $j \geq i$. Hence, for every $i \geq 1$, the alignments up to $\tau_i$ and after $\tau_i$ are independent and $V_{\tau_i+1:\infty}$ agrees with $V_{\tau_1+1:\infty}$ in distribution. Regenerativity of $Z$ with respect to $(\tau_i)_{i \geq 0}$ follows straightforwardly and we refer to the formal proof of [5], Theorem 2, for details.

**Theorem 4.1.** *If $X$ satisfies the assumptions of Lemma 3.1, then there exist probability measures $Q_l(\psi)$, $l \in S$, such that $\hat{Q}_l^n(\psi, X_{1:n}) \Rightarrow_{n \to \infty} Q_l(\psi)$ and $\hat{P}_l^n(\psi, X_{1:n}) \Rightarrow_{n \to \infty} Q_l(\psi)$ almost surely.*

**Proof.** The proof below uses regenerativity of $Z$ in a standard way. For every $n \geq \tau_0$, $A \in \mathcal{B}$ and $l \in S$, we have

$$\frac{1}{n} \sum_{i=1}^{n} I_{A \times \{l\}}(Z_i) = \frac{1}{n} \sum_{i=1}^{\tau_0} I_{A \times \{l\}}(Z_i) + \frac{1}{n} \sum_{i=\tau_0+1}^{\tau_{k(n)}} I_{A \times \{l\}}(Z_i) + \frac{1}{n} \sum_{i=\tau_{k(n)}+1}^{n} I_{A \times \{l\}}(Z_i), \quad (4.3)$$

where $k(n) = \max\{k : \tau_k \leq n\}$ is also a renewal process. Now, since $\tau_0 < \infty$ a.s., we have

$$\frac{1}{n} \sum_{i=1}^{\tau_0} I_{A \times \{l\}}(Z_i) \leq \frac{\tau_0}{n} \xrightarrow[n \to \infty]{} 0, \quad \text{a.s.}$$

Let $\mathcal{M} \stackrel{\text{def}}{=} ET_1$, which is finite by Proposition 4.1. Then, $(n - \tau_{k(n)})/n \leq T_{k(n)+1}/n \to 0$, a.s. Finally, since $Z$ is regenerative with respect to $\tau_0, \tau_1, \ldots$, we have

$$\frac{1}{n} \sum_{i=\tau_0+1}^{\tau_{k(n)}} I_{A \times \{l\}}(Z_i) = \frac{k(n)}{n} \frac{1}{k(n)} \sum_{k=1}^{k(n)} \xi_k, \quad \text{where } \xi_k \stackrel{\text{def}}{=} \sum_{i=\tau_{k-1}+1}^{\tau_k} I_{A \times \{l\}}(Z_i), \ k \geq 1,$$

and are i.i.d. Let $m_l(A; \psi) \stackrel{\text{def}}{=} E\xi_k$. Since $m_l(A; \psi) \leq \mathcal{M} < \infty$, it holds that, as $n \to \infty$,

$$\frac{n}{k(n)} \to \mathcal{M} \quad \text{and} \quad \frac{1}{k(n)} \sum_{k=1}^{k(n)} \xi_k \to m_l(A; \psi) \quad \text{a.s.,}$$

implying that (4.3) tends to $m_l(A;\psi)/\mathcal{M}$ a.s. Similarly,

$$\frac{1}{n}\sum_{i=1}^{n}I_{\{l\}}(V_i) \to \frac{w_l}{\mathcal{M}} \leq 1 \qquad \text{a.s., where } w_l(\psi) \stackrel{\text{def}}{=} E\left(\sum_{i=\tau_{k-1}+1}^{\tau_k}I_{\{l\}}(V_i)\right).$$

Hence, we have shown that for each $l \in S$ and every $A \in \mathcal{B}$,

$$\hat{Q}_l^n(A;Z_{1:n}) \xrightarrow[n\to\infty]{} Q_l(A;\psi), \qquad \text{a.s., where } Q_l(A;\psi) \stackrel{\text{def}}{=} m_l(A;\psi)/w_l.$$

It is easy to note that $A \mapsto m_l(A;\psi)$ is a measure and that $m_l(\mathcal{X};\psi) = w_l(\psi)$. Hence, every $Q_l(\psi)$ ($l \in S$) is a probability measure. Recalling that $\mathcal{X}$ is a separable metric space and invoking the theory of weak convergence of measures now establishes that $\hat{Q}_l^n(Z_{1:n}) \underset{n\to\infty}{\Rightarrow} Q_l(\psi)$ almost surely. It remains to show that for all $l \in S$ and $A \in \mathcal{B}$,

$$\hat{P}_l^n(A;\psi,X_{1:n}) \xrightarrow[n\to\infty]{} Q_l(A;\psi), \qquad \text{a.s.} \tag{4.4}$$

To see this, consider $\sum_{i=1}^{n} I_{A\times\{l\}}(X_i,V_i')$. Since $V_i' = V_i$ for $i \leq \tau_{k(n)}$, we obtain

$$\frac{1}{n}\sum_{i=1}^{n}I_{A\times\{l\}}(X_i,V_i')$$

$$= \frac{1}{n}\sum_{i=1}^{\tau_0}I_{A\times\{l\}}(Z_i) + \frac{1}{n}\sum_{i=\tau_0+1}^{\tau_{k(n)}}I_{A\times\{l\}}(Z_i) + \frac{1}{n}\sum_{i=\tau_{k(n)}+1}^{n}I_{A\times\{l\}}(X_i,V_i')$$

$$\xrightarrow[n\to\infty]{\text{a.s.}} m_l(A;\psi)/\mathcal{M}.$$

Similarly, $\frac{1}{n}\sum_{i=1}^{n}I_{\{l\}}(V_i') \xrightarrow[n\to\infty]{} w_l/\mathcal{M}$ almost surely. $\square$

**Corollary 4.1.** *If $X_{1:\infty}$ satisfies the assumptions of Lemma 3.1, then, for every $l \in S$, there exists a probability measure $q_{l1},\ldots,q_{lK}$ on $S$ such that $\hat{p}_{lj}^n(\psi;X_{1:n}) \xrightarrow[n\to\infty]{} q_{lj}(\psi)$ and $\hat{q}_{lj}^n(Z_{1:n}) \xrightarrow[n\to\infty]{} q_{lj}(\psi)$ almost surely.*

**Proof.** The proof is the same as that of Theorem 4.1, with

$$q_{lj}(\psi) \stackrel{\text{def}}{=} \frac{w_{lj}(\psi)}{w_l(\psi)}, \qquad w_{lj}(\psi) \stackrel{\text{def}}{=} E\left(\sum_{i=\tau_1+1}^{\tau_2}I_{\{l\}}(V_i)I_{\{j\}}(V_{i+1})\right).$$

$\square$

## 5. Conclusion

We have proposed, in [27], [24] and in this work, to improve the precision of the VT estimation by enabling the estimation algorithm to asymptotically confirm the true parameters. In this work, we have developed the central theoretical component of the above methodology. Namely, we have constructed a suitable infinite Viterbi decoding process and have used it to prove the existence of the limiting distributions responsible for the 'fixed point bias' in a very general class of HMMs. General approaches to the efficient computing of the correction functions have been recently proposed in [24]. Model-specific implementations of these approaches are a subject of the authors' continuing investigation.

## Acknowledgements

The first author has been supported by Estonian Science Foundation Grant 5694. The authors are thankful to *EURANDOM* (The Netherlands) and Professors R. Gill and A. van der Vaart for their support. The authors also thank the anonymous referees and Associate Editor for their critical and constructive comments which have helped to improve this manuscript.